\documentclass{revtex4-2}
\usepackage{amsmath,amssymb,bm}
\usepackage{graphicx}
\usepackage{booktabs}
\usepackage[colorlinks=true,linkcolor=blue,citecolor=magenta,urlcolor=blue]{hyperref}

\newcommand{\eps}{\varepsilon}
\newcommand{\thGR}{\theta_{\rm GR}}

\begin{document}

\title{Testing Charged Black Holes Embedded in Anisotropic Matter and Surrounded by Plasma through Gravitational Lensing: Bounds from the Sloan Lens ACS Survey}

\author{Ritesh Pandey}
\email{riteshphy0@gmail.com}
\affiliation{Centre for Space Research, North-West University, Potchefstroom 2520, South Africa}

\author{Shubham Kala}
\email{shubhamkala871@gmail.com}
\affiliation{Centre for Space Research, North-West University, Potchefstroom 2520, South Africa}

\author{Amare Abebe}
\email{Amare.Abebe@nithecs.ac.za}
\affiliation{Centre for Space Research, North-West University, Potchefstroom 2520, South Africa}
\affiliation{National Institute for Theoretical and Computational Sciences (NITheCS), South Africa}

\author{Hemwati Nandan}
\email{hnandan@associates.iucaa.in}
\affiliation{Department of Physics, Hemvati Nandan Bahuguna Garhwal Central University, Srinagar, Garhwal, Uttarakhand 246174, India}
\affiliation{Centre for Space Research, North-West University, Potchefstroom 2520, South Africa}

\author{G. G. L. Nashed}
\email{nashed@bue.edu.eg}
\affiliation{Centre for Theoretical Physics, The British University in Egypt, Cairo 11837, Egypt}
\affiliation{Centre for Space Research, North-West University, Potchefstroom 2520, South Africa}

\begin{abstract}
We investigate weak gravitational lensing by a charged black hole surrounded by anisotropic matter, described by the two-component Kiselev metric, and immersed in a homogeneous plasma. We first examine the black hole geometry and horizon structure and then derive the weak-deflection angle from the Hamiltonian description of light propagation in plasma, using an expansion with a fixed turning point that we verify against direct numerical integration. Using the gravitational lens equation, we obtain the Einstein-ring angle, the image positions, the total magnification and the fractional magnifications of the primary and secondary images. We then confront the model with the Einstein radii of two galaxy-scale lenses of the Sloan Lens ACS Survey, J0008$-$0004 and J0029$-$0055. Because the Einstein angle cannot be smaller than the reduced deficit deflection, the observed radii give the mass-independent bounds $K<6.5\times10^{-6}$ and $K<4.2\times10^{-6}$, respectively. The charge correction to the Einstein radius is suppressed by $r_s/R_E\approx 5\times10^{-6}$ and is unobservable, while the plasma parameter at the optical wavelengths of the SLACS imaging is of order $10^{-22}$ and cannot be constrained. Our results show that galaxy-scale strong lensing data place stringent limits on anisotropic matter surrounding the lens, whereas plasma effects can be probed only at low radio frequencies.
\\

\end{abstract}

\maketitle

\section{Introduction}

Black holes~(BHs) are among the most intriguing and compact objects predicted by general relativity~\cite{einstein1916,schwarzschild1916,hawking1972,weinberg1972}. Since their theoretical prediction, they have attracted extensive attention in both theoretical and observational studies, and recent observations have strengthened their role as natural laboratories for testing the predictions of strong-field gravity. The extreme gravitational environment surrounding BHs provides a unique setting for investigating a wide range of astrophysical and fundamental phenomena~\cite{berti2015}. Their strong gravitational fields, event horizons, spacetime curvature and frame-dragging effects offer valuable opportunities to explore the behaviour of matter, radiation and spacetime under extreme conditions~\cite{barack2019}. Consequently, BHs have become important theoretical tools for studying the fundamental nature of gravity and several unresolved phenomena in the Universe~\cite{yunes2016}.

Observational investigations of BHs have progressed substantially over the past decade. The first direct detection of gravitational waves from a binary BH merger, GW150914, provided evidence for the existence of stellar-mass binary BH systems~\cite{LIGOScientific:2016lio}. Subsequently, the Event Horizon Telescope (EHT) collaboration obtained horizon-scale images of the compact object in the centre of the galaxy M87, revealing a bright emission ring surrounding a central dark region consistent with the expected BH shadow~\cite{eht2019}. The EHT observation of Sgr~A$^*$ subsequently provided horizon-scale observations of the compact object at the centre of the Milky Way~\cite{eht2022}. These observations have opened new avenues for investigating BH spacetimes through both gravitational and electromagnetic observations.

Astrophysical BHs are expected to interact with their surrounding media rather than exist as completely isolated objects. Matter and energy in the vicinity of a BH can take various forms, including anisotropic fluids~\cite{papadopoulos1999}, perfect fluids~\cite{villas2000}, dark matter distributions~\cite{rizwan2019}, homogeneous and inhomogeneous plasma~\cite{bk2015} and string clouds~\cite{letelier1979}. The presence of these components can alter the spacetime geometry and, consequently, the properties of the BH. Such modifications may influence the structure of the event and Cauchy horizons, the geodesic motion of particles and photons, and the thermodynamic behaviour of the system. Studying BHs in different surrounding environments therefore provides a broader framework for understanding their physical properties and their possible astrophysical manifestations. Recently, Garzon \textit{et al.}~\cite{garzon2026} modelled the environment of a charged BH as matter with anisotropic pressure, using the Kiselev family of solutions~\cite{kiselev2003}, and analysed its horizon structure and quasinormal-mode spectrum. Kiselev spacetimes are static, spherically symmetric solutions of the Einstein equations whose source is an effective stress-energy tensor with unequal radial and tangential pressures. Visser~\cite{visser2020} showed that this tensor is in general neither a perfect fluid nor quintessence in the cosmological sense, and it has been shown that it can be split into a perfect-fluid part and an electromagnetic part~\cite{qu2023}. For the equation-of-state parameter $w=-1/3$ considered here, the Kiselev term is a constant shift of the lapse function, which is geometrically equivalent to a cloud of strings~\cite{letelier1979} or a global monopole~\cite{barriola1989}; the resulting spacetime is not asymptotically flat but carries a solid-angle deficit. As we show below, this deficit leaves a distinctive imprint on light deflection.

The surrounding medium need not be limited to an anisotropic fluid; a plasma environment can also be incorporated into the description of light propagation in the BH spacetime. The presence of plasma gives the medium a non-trivial refractive index, causing photons to propagate differently from their trajectories in vacuum~\cite{ehlers1987,breuer1982}. The refractive index depends on the photon frequency and on the local plasma properties, which makes the optical behaviour of the spacetime dispersive~\cite{synge1960,rogers2015}. This modification of photon propagation can be incorporated through a Hamiltonian formulation, which provides a convenient framework for describing light rays in a dispersive plasma. Bisnovatyi-Kogan and Tsupko showed that even a homogeneous plasma modifies the gravitational bending angle and makes it frequency dependent~\cite{bk2010,bk2015}. For stationary and axisymmetric spacetimes, the Hamilton--Jacobi equation can be separated under specific conditions on the metric and the plasma-frequency distribution, allowing the photon region and shadow to be determined analytically~\cite{bezdekova2022}. In the particular case of the Kerr spacetime, separable plasma distributions admit a Carter-like constant, and the presence of plasma leads to several features absent in vacuum, including modified spherical photon orbits, off-equatorial circular photon orbits and non-monotonic behaviour of the light-deflection angle~\cite{perlick2024}. The influence of plasma has since been examined in various BH spacetimes, showing that the properties of the surrounding medium can modify photon trajectories and lensing observables~\cite{morozova2013,babar2021,kala2022,kala2024,kala2025,khasanov2025,mushtaq2025a,mushtaq2025b,roy2026,pandey2026a,ovgun2026,das2026,kala2026a}.

The weak-deflection regime provides a perturbative framework for studying gravitational lensing when the photon impact parameter is much larger than the characteristic scale of the BH~\cite{acquaviva2004,keeton2005,amore2007,iyer2007,jia2020}. The deflection angle can then be incorporated into the lens equation to determine observables such as the Einstein-ring radius, image positions, angular separation and magnification~\cite{virbhadra2000,frittelli2000,bozza2002,eiroa2011,bartelmann2010}. Weak gravitational lensing in the presence of electric charge, dark matter, anisotropic matter and plasma has been investigated in several studies. Atamurotov \textit{et al.}~\cite{atamurotov2021} investigated the influence of axion--plasmon coupling on photon motion around a Schwarzschild BH and demonstrated its effects on the BH shadow, the gravitational deflection angle and Einstein rings. Their results highlight the sensitivity of lensing observables to the properties of the surrounding plasma. Magnetic fields can also have a significant impact on the optical structure of BHs: Lima Junior \textit{et al.}~\cite{lima2021} demonstrated that strong magnetic fields modify null geodesics, photon orbits, shadow morphology and lensing features. Furthermore, Hoshimov \textit{et al.}~\cite{hoshimov2024} investigated weak gravitational lensing and BH shadows in a GUP-modified Schwarzschild spacetime surrounded by plasma and showed that plasma and GUP corrections modify the shadow radius, deflection angle, image positions and magnification. Weak gravitational lensing has also been investigated extensively in modified gravity and in other BH spacetimes~\cite{kala2020,ditta2023,albadawi2024,gutierrez2025,yildiz2024,xie2026,ibrokhimov2025,sucu2025,pantig2026,pandey2026b,kala2026b,ahmed2026}.

The EHT collaboration has provided direct observational evidence of compact, ring-like structures associated with the supermassive BHs in M87 and Sgr~A$^*$. The measured angular diameters of the emission rings, $42\pm3~\mu$as for M87 and $51.8\pm2.3~\mu$as for Sgr~A$^*$, provide observational benchmarks for theoretical BH models~\cite{eht2019,eht2022} and have been widely used to constrain modified and non-Kerr BH spacetimes~\cite{vagnozzi2023}. Complementary constraints can be obtained from gravitational lensing. The Sloan Lens ACS (SLACS) Survey provides a large, homogeneous sample of galaxy-scale strong-lensing systems with measured Einstein radii and spectroscopic redshifts~\cite{bolton2006,koopmans2006,bolton2008}. Rahmatov \textit{et al.}~\cite{rahmatov2026} studied weak lensing and microlensing of charged BHs in scalar--tensor--vector gravity and used the SLACS lenses J0008$-$0004 and J0029$-$0055 to constrain the charge and MOG parameters.

Motivated by these developments, we investigate weak gravitational lensing by a charged BH surrounded by anisotropic matter and immersed in a homogeneous plasma. We derive the deflection angle with an expansion that keeps the turning point fixed, verify it against direct numerical integration, and examine how the electric charge, the anisotropic matter and the plasma modify the lensing observables. We then use the Einstein radii of two SLACS lenses to bound the model parameters, using only information that does not already assume general relativity.

The manuscript is organised as follows. Section~\ref{sec:metric} presents the BH metric and horizon structure. Section~\ref{sec:deflection} derives the weak-deflection angle in a homogeneous plasma. Section~\ref{sec:lens} describes the lens geometry and image formation. Section~\ref{sec:slacs} derives bounds from the SLACS Einstein radii. Finally, Sec.~\ref{sec:conclusion} summarises the main results. For the theoretical analysis, we adopt geometrized units, $G=M=c=1$. However, when bound the model parameters using observational data, we use the physical units and parameter conventions adopted in the SLACS dataset.

\section{Black hole metric and horizon structure}\label{sec:metric}
To investigate the effects of an anisotropic environment on the phenomena considered in this work, we adopt the BH spacetime derived in Ref.~\cite{garzon2026}. This solution is based on the Kiselev metric, which describes a static, spherically symmetric spacetime surrounded by additional matter or energy components. The metric is written as
\begin{equation}
ds^2=-f(r)\,dt^2+\frac{dr^2}{f(r)}
+r^2d\Omega^2,
\label{eq:metric}
\end{equation}
where $d\Omega^2=d\theta^2+\sin^2\theta\,d\phi^2$ and the metric function takes the general form~\cite{kiselev2003}
\begin{equation}
f(r)=1-\frac{2M}{r}
-\sum_i\frac{K_i}{r^{3w_i+1}}.
\label{eq:kiselev}
\end{equation}
Here, $M$ denotes the BH mass, while $K_i$ and $w_i$ characterize the strength and equation-of-state parameter of the $i$-th matter or energy component, respectively. The contributions of distinct components are assumed to be additive and non-interacting within this framework.

Following Ref.~\cite{garzon2026}, we consider a two-component model describing a charged BH surrounded by an anisotropic fluid. The first component corresponds to the electromagnetic contribution, specified by
\begin{equation}
K_1=-Q^2,\qquad w_1=\frac{1}{3},
\end{equation}
where $Q$ is the BH charge parameter. The second component represents the surrounding anisotropic fluid and is characterized by
\begin{equation}
K_2=K,\qquad w_2=-\frac{1}{3},
\end{equation}
where $K$ determines the strength of the environmental contribution. 
Substituting these parameters into Eq.~\eqref{eq:kiselev}, the metric function becomes
\begin{equation}
f(r)=1-\frac{2M}{r}+\frac{Q^2}{r^2}-K.
\label{eq:lapse}
\end{equation}
To simplify the analysis, we used the dimensionless coordinates and parameters by rescaling the relevant quantities with the Schwarzschild radius, $r_g=2M$, as
\begin{equation}
r_{\mathrm{an}}=\frac{r}{r_g},\qquad
t_{\mathrm{an}}=\frac{t}{r_g},\qquad
ds_{\mathrm{an}}^2=\frac{ds^2}{r_g^2},
\end{equation}
and
\begin{equation}
Q_{\mathrm{an}}=\frac{Q}{r_g},\qquad
r_{g,\mathrm{an}}=\frac{r_g}{r_g}=1.
\end{equation}
Here, $r$, $t$, $ds^2$, and $Q$ denote the corresponding dimensional quantities. Henceforth, we omit the subscript ``an (anisotropic)'' for notational simplicity.\\

\begin{figure}[t]
\centering
\includegraphics[width=0.5\textwidth]{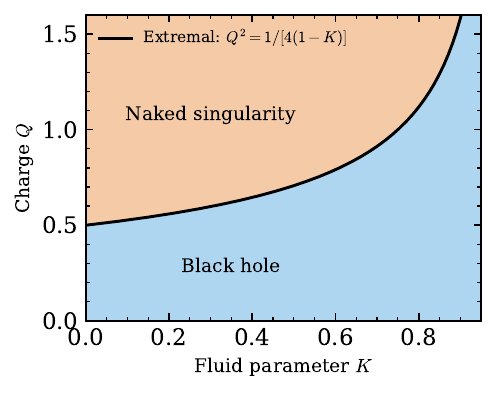}
\caption{Parameter space in the $(K,Q)$ plane. The solid curve is the extremality condition $Q^2=1/[4(1-K)]$. Below it the metric~\eqref{eq:lapse} has an event horizon; above it no horizon exists.}
\label{fig:params}
\end{figure}

In the limit $K=0$, Eq.~\eqref{eq:lapse} reduces to the Reissner--Nordstr\"om (RN)-like solution, while for $Q=0$ it describes a Schwarzschild-like geometry with a solid-angle deficit governed by $K$: for $r\to\infty$, $f\to1-K$, and the rescaling $r=(1-K)^{1/2}\rho$, $t=(1-K)^{-1/2}\tau$ brings the asymptotic metric to the form $-d\tau^2+d\rho^2+(1-K)\rho^2d\Omega^2$. The horizons are the roots of $f(r_\pm)=0$,
\begin{equation}
r_\pm=\frac{1\pm\sqrt{1-4(1-K)Q^2}}{2(1-K)}, \label{eq:horizons}
\end{equation}
where $r_+$ and $r_-$ denote the outer (event) and inner (Cauchy) horizons, respectively. For a BH to exist we require $0\le K<1$ and a non-negative discriminant, $Q^2\le1/[4(1-K)]$, which generalises the extremality bound of the RN spacetime. The corresponding parameter space is shown in Fig.~\ref{fig:params}. As $K$ increases towards unity, the event horizon shifts to larger radii, with $r_+\to1/(1-K)$ for $Q\to0$, reflecting the growing gravitational influence of the surrounding anisotropic matter.

\section{Weak gravitational lensing in a homogeneous plasma}\label{sec:deflection}

The propagation of light in a homogeneous, non-magnetised plasma is described by the Hamiltonian formalism (see Refs.~\cite{bk2015,perlick2024} and references therein). For the static, spherically symmetric spacetime~\eqref{eq:metric}, the Hamiltonian constraint reads
\begin{equation}
\mathcal{H}=\frac12\left(g^{\mu\nu}p_\mu p_\nu+\omega_p^2\right)=0, \label{eq:hamiltonian}
\end{equation}
where $\omega_p$ denotes the constant electron plasma frequency. Restricting the photon motion to the equatorial plane ($\theta=\pi/2$), the conserved quantities associated with the timelike and axial Killing vectors are
\begin{equation}
p_t=-\omega_0,\qquad p_\varphi=L, \label{eq:conserved}
\end{equation}
where $\omega_0$ and $L$ are the conserved photon energy and angular momentum. Eq.~\eqref{eq:hamiltonian} then becomes
\begin{equation}
-\frac{\omega_0^2}{f(r)}+f(r)p_r^2+\frac{L^2}{r^2}+\omega_p^2=0. \label{eq:constraint}
\end{equation}
Using Hamilton's equations, $\dot r=f(r)p_r$ and $\dot\varphi=L/r^2$, the orbit equation is
\begin{equation}
\left(\frac{dr}{d\varphi}\right)^2=\frac{r^4f(r)}{L^2}\left[\frac{\omega_0^2}{f(r)}-\frac{L^2}{r^2}-\omega_p^2\right]. \label{eq:orbit_r}
\end{equation}
Introducing the inverse radial coordinate $u=1/r$, the impact parameter $b=L/\omega_0$ and the plasma parameter
\begin{equation}
\eps\equiv\frac{\omega_p^2}{\omega_0^2}, \label{eq:eps}
\end{equation}
Eq.~\eqref{eq:orbit_r} takes the form $(du/d\varphi)^2=\Phi(u)$ with
\begin{align}
\Phi(u)&=\frac{1}{b^2}-f(u)\left(u^2+\frac{\eps}{b^2}\right)\nonumber\\
&=\frac{1-A\eps}{b^2}+\frac{\eps}{b^2}u-\left(A+\frac{\eps Q^2}{b^2}\right)u^2+u^3-Q^2u^4, \label{eq:Phi}
\end{align}
where
\begin{equation}
A\equiv1-K. \label{eq:A}
\end{equation}
Since $f\to A$ at large $r$, the frequency measured by a static observer far from the hole is $\omega_0/\sqrt{A}$, and the corresponding local plasma parameter there is $A\eps$; this combination appears in all results below. The deflection angle is obtained from~\cite{weinberg1972,keeton2005}
\begin{equation}
\hat\alpha+\pi=2\int_0^{u_0}\frac{du}{\sqrt{\Phi(u)}}, \label{eq:deflection_integral}
\end{equation}
where the turning point $u_0$ satisfies $\Phi(u_0)=0$.\\

In the weak-field regime $b\gg1$ (that is, $b\gg2M$), $u\le u_0\ll1$ along the ray. A direct expansion of the integrand of Eq.~\eqref{eq:deflection_integral} in the $u^3$ and $u^4$ terms with the upper limit held fixed is not permissible: the turning points of the full and truncated $\Phi$ differ, and the resulting integrands diverge as $(u_0-u)^{-3/2}$ at the end point. We therefore eliminate $b$ in favour of $u_0$ using $\Phi(u_0)=0$,
\begin{equation}
\frac{1}{b^2}=\frac{f(u_0)\,u_0^2}{1-\eps f(u_0)}, \label{eq:b_u0}
\end{equation}
and introduce $x=u/u_0$, which fixes the end point at $x=1$~\cite{keeton2005}. Eq.~\eqref{eq:deflection_integral} becomes
\begin{align}
\hat\alpha+\pi&=2\int_0^1\frac{dx}{\sqrt{G(x)}},\nonumber\\
G(x)&=\frac{f(u_0)\left[1-\eps f(u_0x)\right]}{1-\eps f(u_0)}-x^2f(u_0x). \label{eq:G}
\end{align}
At zeroth order in $u_0$, $G=A(1-x^2)$ and
\begin{equation}
\hat\alpha^{(0)}=\pi\left[(1-K)^{-1/2}-1\right]\simeq\frac{\pi K}{2}, \label{eq:alpha0}
\end{equation}
which is independent of $b$ and of $\eps$ and vanishes for $K=0$. This is the deflection associated with the solid-angle deficit and coincides with that of a global monopole or a cloud of strings~\cite{barriola1989}. At first order in $u_0$,
\begin{equation}
G(x)\simeq A(1-x^2)-u_0\left[(1-x^3)+\frac{A\eps(1-x)}{1-A\eps}\right],
\end{equation}
and, with
\begin{equation}
\int_0^1\frac{1-x^3}{(1-x^2)^{3/2}}dx=2,\qquad\int_0^1\frac{1-x}{(1-x^2)^{3/2}}dx=1, \label{eq:integrals}
\end{equation}
we find
\begin{equation}
\hat\alpha=\hat\alpha^{(0)}+\frac{u_0}{A^{3/2}}\left[2+\frac{A\eps}{1-A\eps}\right]+\mathcal{O}(u_0^2). \label{eq:alpha_u0}
\end{equation}
Inverting Eq.~\eqref{eq:b_u0} at leading order, $u_0=\sqrt{1-A\eps}/(\sqrt{A}\,b)+\mathcal{O}(b^{-2})$, gives
\begin{equation}
\hat\alpha=\hat\alpha^{(0)}+\frac{2-A\eps}{A^2\,b\sqrt{1-A\eps}}+\mathcal{O}(b^{-2}). \label{eq:alpha_first}
\end{equation}

At second order we retain only the vacuum terms, because cross terms of order $\eps/b^2$ lie beyond the accuracy required for the applications below. For $\eps=0$, the substitution $r=A^{1/2}\rho$, $t=A^{-1/2}\tau$, $\varphi=A^{-1/2}\varphi'$ maps the metric~\eqref{eq:metric} exactly onto the RN metric with mass $M_{\rm eff}=MA^{-3/2}$ and charge $Q_{\rm eff}=QA^{-1}$, while the impact parameter $L/\omega_0$ is unchanged. The standard second-order RN result~\cite{keeton2005} for the deflection in $\varphi'$, divided by $A^{1/2}$, then gives

\begin{align}
\hat\alpha={}&\pi\left[A^{-1/2}-1\right]+\frac{2-A\eps}{A^2\,b\sqrt{1-A\eps}} +\frac{15\pi}{16A^{7/2}b^2}-\frac{3\pi Q^2}{4A^{5/2}b^2}+\mathcal{O}\!\left(\frac{\eps}{b^2},\frac{1}{b^3}\right). \label{eq:alpha_full}
\end{align}

\begin{figure*}[t]
\centering
\includegraphics[width=0.92\textwidth]{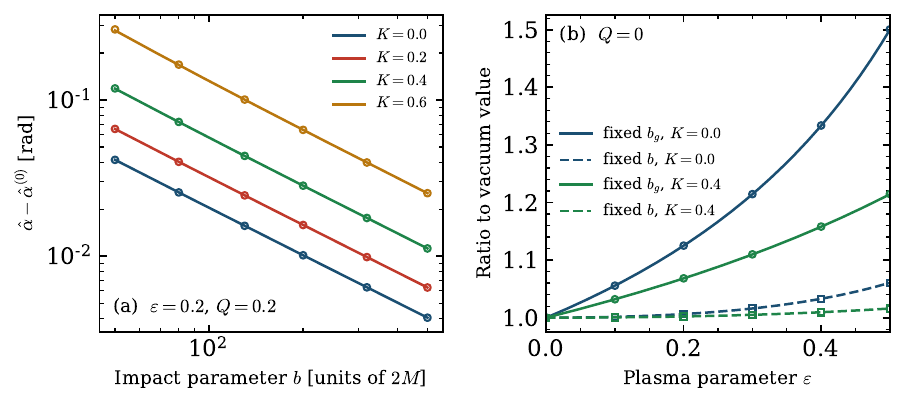}
\caption{(a) The impact parameter dependent part of deflection angle, as a function of the impact parameter for different values of $K$, with $\eps=0.2$ and $Q=0.2$. Curves: Eq.~\eqref{eq:alpha_full}; open circles: numerical integration of Eq.~\eqref{eq:deflection_integral}. (b) Ratio of $\hat\alpha-\hat\alpha^{(0)}$ to its vacuum value as a function of $\eps$ for $K=0$ and $0.4$ and $Q=0$, at fixed geometric impact parameter $b_g$ (solid curves, circles) and at fixed $b$ (dashed curves, squares), evaluated at $b_g$ or $b=1000$.}
\label{fig:deflection}
\end{figure*}

\begin{figure*}[t]
\centering
\includegraphics[width=0.92\textwidth]{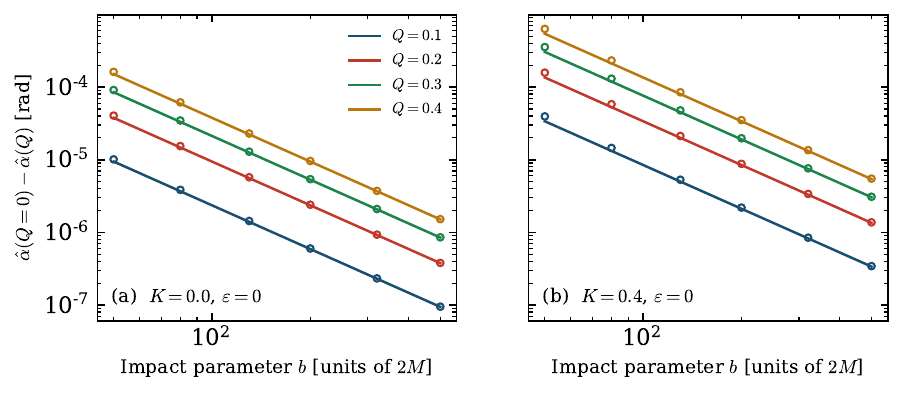}
\caption{Reduction of the deflection angle by the charge, $\hat\alpha(Q=0)-\hat\alpha(Q)$, as a function of $b$ for $\eps=0$ and (a) $K=0$, (b) $K=0.4$. Curves: the $Q^2$ term of Eq.~\eqref{eq:alpha_full}; open circles: numerical integration.}
\label{fig:charge}
\end{figure*}

The parameter $b=L/\omega_0$ is not the quantity tied directly to the observed image position. For a static observer at areal radius $r_O$, the angle $\theta$ between the incoming ray and the radial direction satisfies
\begin{equation}
\sin\theta=\frac{b\sqrt{f(r_O)}}{r_O\sqrt{1-\eps f(r_O)}}, \label{eq:sin_theta}
\end{equation}
because the magnitude of the local photon momentum is $(\omega_0^2/f-\omega_p^2)^{1/2}$. Far from the lens, $\theta\simeq b_g/r_O$, with
\begin{equation}
b_g=\frac{b\sqrt{A}}{\sqrt{1-A\eps}} \label{eq:bg}
\end{equation}
the geometric impact parameter of the ray~\cite{bk2010}.
In terms of $b_g$, Eq.~\eqref{eq:alpha_full} reads
\begin{equation}
\hat\alpha=\hat\alpha^{(0)}+\frac{c}{b_g}+\frac{e}{b_g^2}, \label{eq:alpha_bg}
\end{equation}
with
\begin{equation}
c=A^{-3/2}\left[1+\frac{1}{1-A\eps}\right],\qquad e=\frac{15\pi}{16A^{5/2}}-\frac{3\pi Q^2}{4A^{3/2}}, \label{eq:ce}
\end{equation}
where $e$ is evaluated at $\eps=0$. To leading order in $K$, $\eps$ and $Q^2$,
\begin{equation}
\hat\alpha\simeq\frac{\pi K}{2}+\frac{2}{b_g}\left(1+\frac{3K}{2}+\frac{\eps}{2}\right)+\frac{1}{b_g^2}\left(\frac{15\pi}{16}-\frac{3\pi Q^2}{4}\right). \label{eq:alpha_leading}
\end{equation}

Eqs.~\eqref{eq:alpha_bg}--\eqref{eq:alpha_leading} show three distinct effects. First, $K$ produces the constant deflection~\eqref{eq:alpha0} together with an enhancement of the $1/b_g$ term, equivalent to an effective mass $M(1-K)^{-3/2}$; the surrounding anisotropic matter therefore increases, rather than screens, the bending of light. Second, the charge enters only at second order and with a negative sign, exactly as in the RN spacetime, so that $Q$ weakens the deflection. Third, at fixed $b_g$ the homogeneous plasma increases the deflection by the factor $[1+1/(1-A\eps)]/2\simeq1+\eps/2$, and for $K=0$ Eq.~\eqref{eq:alpha_bg} reduces to the result of Bisnovatyi-Kogan and Tsupko~\cite{bk2010}. The plasma correction vanishes as $b_g\to\infty$, as it must, since a homogeneous medium does not bend light in flat spacetime. At fixed $b$, by contrast, the plasma correction to the first-order term is of order $\eps^2$ [Eq.~\eqref{eq:alpha_first}], so statements about the plasma effect depend on which impact parameter is held fixed; only $b_g$ is directly related to observables.

We verified Eq.~\eqref{eq:alpha_full} by evaluating Eq.~\eqref{eq:deflection_integral} numerically at 30-digit precision, using the substitution $u=u_0(1-s^2)$ to remove the end-point singularity. Table~\ref{tab:validation} compares the two for representative parameters. The $b$-dependent part of the deflection agrees to a few parts in $10^4$ at $b=100$, consistent with the neglected $\mathcal{O}(b^{-3})$ and $\mathcal{O}(\eps/b^2)$ terms, and the agreement improves as $b$ increases.

\begin{table*}[t]
\caption{Comparison of Eq.~\eqref{eq:alpha_full} with direct numerical integration of Eq.~\eqref{eq:deflection_integral}. Lengths are in units of $2M$; the last three columns give $\hat\alpha-\hat\alpha^{(0)}$ and the relative difference.}
\label{tab:validation}
\begin{ruledtabular}
\begin{tabular}{rcccccc c}
$b$ & $K$ & $\eps$ & $Q$ & $\hat\alpha^{(0)}$ & numerical & Eq.~\eqref{eq:alpha_full} & rel.\ diff.\\
\hline
100 & 0 & 0 & 0 & 0 & $2.03000\times10^{-2}$ & $2.02945\times10^{-2}$ & $2.7\times10^{-4}$\\
100 & 0 & 0 & 0.4 & 0 & $2.02610\times10^{-2}$ & $2.02568\times10^{-2}$ & $2.0\times10^{-4}$\\
100 & 0.2 & 0 & 0.3 & 0.37082 & $3.18710\times10^{-2}$ & $3.18561\times10^{-2}$ & $4.7\times10^{-4}$\\
100 & 0 & 0.2 & 0 & 0 & $2.04124\times10^{-2}$ & $2.04191\times10^{-2}$ & $3.3\times10^{-4}$\\
300 & 0.2 & 0.2 & 0.3 & 0.37082 & $1.05221\times10^{-2}$ & $1.05236\times10^{-2}$ & $1.5\times10^{-4}$\\
1000 & 0.4 & 0.3 & 0.2 & 0.91419 & $5.59964\times10^{-3}$ & $5.60019\times10^{-3}$ & $9.8\times10^{-5}$
\end{tabular}
\end{ruledtabular}
\end{table*}

Fig.~\ref{fig:deflection}(a) shows the $b$-dependent part of the deflection for several values of $K$ at $\eps=0.2$ and $Q=0.2$, together with the numerical integration (open circles). The deflection decreases as $1/b$ and increases with $K$ through the factor $(1-K)^{-2}$ in Eq.~\eqref{eq:alpha_full}. The constant term, which is not plotted, equals $\hat\alpha^{(0)}=0.371$, $0.914$ and $1.83$~rad for $K=0.2$, $0.4$ and $0.6$; such values of $K$ are shown only to illustrate the scaling, since Sec.~\ref{sec:slacs} shows that $K\lesssim10^{-5}$ for galaxy-scale lenses. Fig.~\ref{fig:deflection}(b) shows the plasma dependence at fixed $b_g$ (solid curves) and at fixed $b$ (dashed curves), normalised to the vacuum value: the former follows $[1+1/(1-A\eps)]/2$, while the latter changes only at order $\eps^2$.

Fig.~\ref{fig:charge} isolates the charge contribution, $\hat\alpha(Q=0)-\hat\alpha(Q)$, for $\eps=0$ and $K=0$ and $0.4$. It is positive, that is, the charge reduces the deflection; it scales as $Q^2/b^2$ and grows with $K$ through the factor $(1-K)^{-5/2}$, and the numerical results agree with the analytic term. At $b=100$ and $Q=0.4$ the charge changes the deflection by $3.8\times10^{-5}$~rad, more than two orders of magnitude below the leading term.

\section{Lens geometry and image formation}\label{sec:lens}

\begin{figure}[t]
\centering
\includegraphics[width=0.8\textwidth]{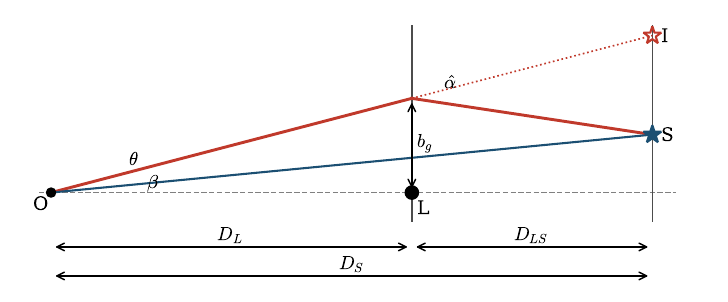}
\caption{Lens geometry: observer O, lens L, source S and image I, with the source angle $\beta$, image angle $\theta$, deflection angle $\hat\alpha$, geometric impact parameter $b_g$ and the angular-diameter distances $D_L$, $D_S$ and $D_{LS}$~\cite{virbhadra2000}.}
\label{fig:geometry}
\end{figure}

Having derived the weak-field deflection angle~\eqref{eq:alpha_bg}, we now employ the thin-lens geometry to obtain the Einstein-ring angle, the image positions and the magnifications. The source, lens and observer are configured as shown in Fig.~\ref{fig:geometry}.

\begin{figure*}[t]
\centering
\includegraphics[width=0.85\textwidth]{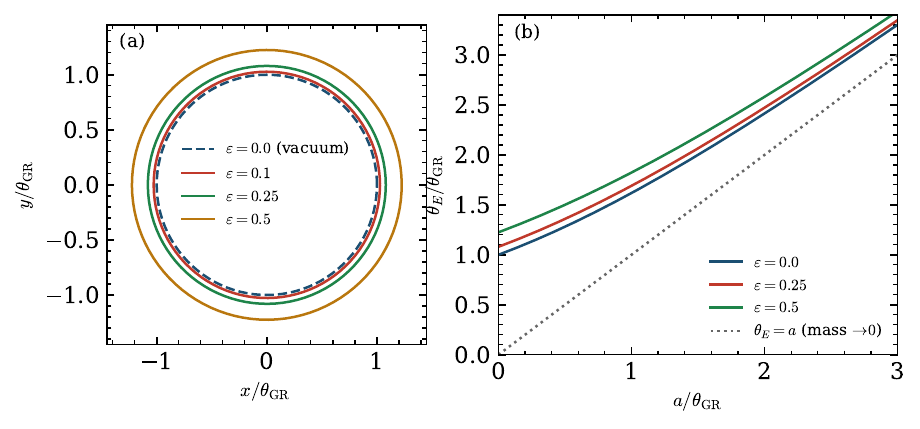}
\caption{(a) Einstein rings for $K=Q=0$ and different values of the plasma parameter $\eps$, normalised to the vacuum Einstein angle $\thGR$ (dashed). (b) Einstein angle as a function of the reduced deficit deflection $a$ [Eq.~\eqref{eq:lens_params}] for different $\eps$, with $K\ll1$ in $\theta_c$; the dotted line $\theta_E=a$ is the limit of vanishing mass.}
\label{fig:ring}
\end{figure*}

\begin{figure*}[t]
\centering
\includegraphics[width=0.92\textwidth]{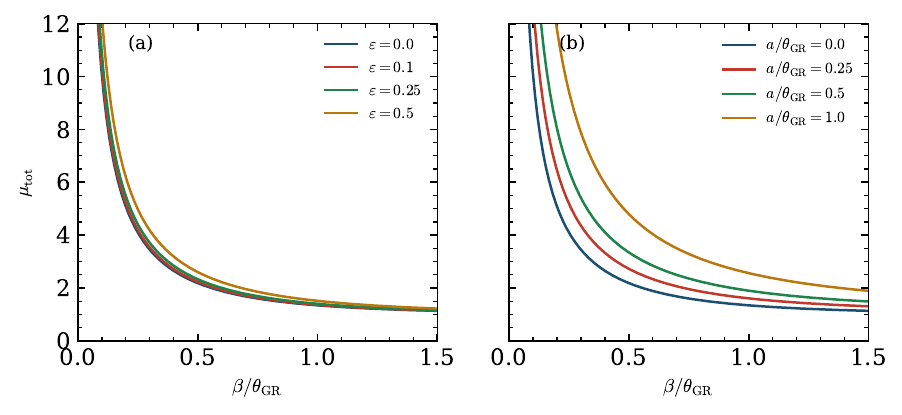}
\caption{Total magnification as a function of the source position normalised to the vacuum Einstein angle, $\beta/\thGR$, with $K\ll1$ in $\theta_c$. (a) $a=0$ and different values of $\eps$; (b) $\eps=0$ and different values of the reduced deficit deflection $a$.}
\label{fig:mutot}
\end{figure*}

\begin{figure*}[t]
\centering
\includegraphics[width=0.92\textwidth]{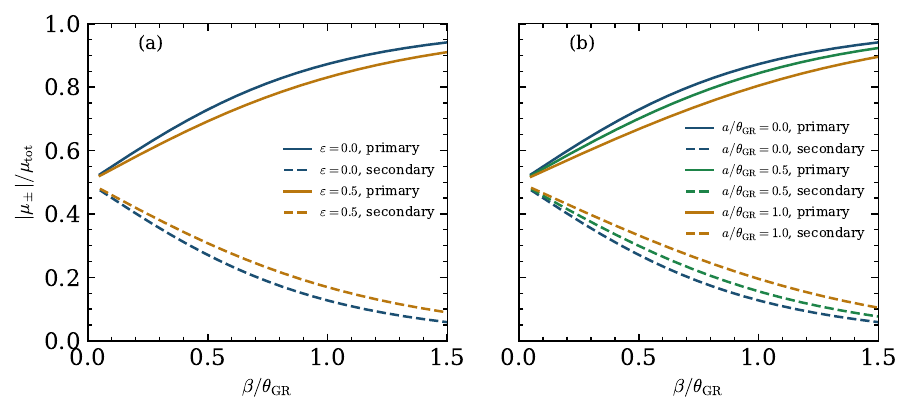}
\caption{Fractional magnifications $|\mu_\pm|/\mu_{\rm tot}$ of the primary (solid) and secondary (dashed) images as functions of $\beta/\thGR$. (a) $a=0$ and $\eps=0$, $0.5$; (b) $\eps=0$ and $a/\thGR=0$, $0.5$, $1$.}
\label{fig:mufrac}
\end{figure*}

\subsection{Lens equation}

In the thin-lens approximation, valid when the deflection occurs over a region much smaller than the source--lens and lens--observer distances, the lens equation relates the angular positions of the source $\beta$ and the image $\theta$ through~\cite{bozza2008}
\begin{equation}
\beta=\theta-\frac{D_{LS}}{D_S}\,\hat\alpha, \label{eq:lens_eq}
\end{equation}
where $D_L$, $D_S$ and $D_{LS}$ are the angular-diameter distances from the observer to the lens, from the observer to the source and from the lens to the source. For a ray with geometric impact parameter $b_g$, Eq.~\eqref{eq:sin_theta} gives $b_g=D_L|\theta|$ at leading order, with $D_L$ identified with the areal distance of the observer from the lens. In the conical region the areal and proper radial distances differ by the factor $(1-K)^{1/2}$; this difference is immaterial for $K\ll1$, which is the only observationally relevant regime (Sec.~\ref{sec:slacs}). Since the deflection always points towards the lens, substituting Eq.~\eqref{eq:alpha_bg} gives
\begin{equation}
\beta=\theta-{\rm sgn}(\theta)\left(a+\frac{\eta}{\theta^2}\right)-\frac{\theta_c^2}{\theta}, \label{eq:lens_eq_full}
\end{equation}
with
\begin{equation}
a\equiv\frac{D_{LS}}{D_S}\hat\alpha^{(0)},\quad\theta_c^2\equiv\frac{c\,r_s D_{LS}}{D_LD_S},\quad\eta\equiv\frac{e\,r_s^2D_{LS}}{D_L^2D_S}, \label{eq:lens_params}
\end{equation}
where we have restored $r_s$. In the Schwarzschild limit $a=0$ and $\theta_c=\thGR$, with $\thGR^2\equiv2r_sD_{LS}/(D_LD_S)$.

\subsection{Einstein ring}

The Einstein ring forms when the source, lens and observer are aligned ($\beta=0$). Solving Eq.~\eqref{eq:lens_eq_full} perturbatively in $\eta$ gives
\begin{equation}
\theta_E=\theta_E^{(1)}+\frac{\eta}{\big(\theta_E^{(1)}\big)^2+\theta_c^2},\qquad\theta_E^{(1)}=\frac{a}{2}+\sqrt{\frac{a^2}{4}+\theta_c^2}. \label{eq:thetaE}
\end{equation}
For $a=0$ this becomes
\begin{align}
\theta_E&\simeq\theta_c\left(1+\frac{e\,r_s}{2c\,R_E}\right),\nonumber\\
\frac{\theta_c}{\thGR}&=(1-K)^{-3/4}\left[\frac{1}{2}+\frac{1}{2[1-(1-K)\eps]}\right]^{1/2}, \label{eq:thetaE_a0}
\end{align}
where $R_E=D_L\theta_E$ is the physical Einstein radius at the lens plane~\cite{younas2015}.

Several physical observations follow from Eqs.~\eqref{eq:thetaE} and~\eqref{eq:thetaE_a0}. The fluid parameter increases $\theta_E$ through both the effective mass contribution to $\theta_c$ and, more strongly, through $a$. Since $\theta_E>a$ for any positive mass, the deficit term sets a lower bound on the ring size that is independent of $M$, which we exploit in Sec.~\ref{sec:slacs}. The plasma also enlarges the ring; for $K=0$, $\theta_E^2/\thGR^2=(2-\eps)/[2(1-\eps)]\simeq1+\eps/2$. The charge reduces the ring by the fractional amount $(3\pi/16)Q^2r_s/R_E$ for $K,\eps\to0$, with this correction suppressed by $r_s/R_E$, whereas the second-order GR term increases it by $(15\pi/64)r_s/R_E$. In the limit $K=Q=\eps=0$, Eq.~\eqref{eq:thetaE} reduces to the Schwarzschild Einstein angle including its second-order correction.

Fig.~\ref{fig:ring}(a) shows Einstein rings for $K=Q=0$ and increasing $\eps$, normalized to $\thGR$; the ring grows by 3\%, 8\% and 22\% for $\eps=0.1$, $0.25$ and $0.5$. Fig.~\ref{fig:ring}(b) shows $\theta_E$ as a function of the reduced deficit deflection $a$: for $a\gg\theta_c$ the ring follows $\theta_E\simeq a$, independently of the mass and of the plasma.

\subsection{Image positions}

For a source at $\beta>0$, neglecting the small $\eta$ term, Eq.~\eqref{eq:lens_eq_full} admits two images,
\begin{align}
\theta_+&=\frac12\left[(\beta+a)+\sqrt{(\beta+a)^2+4\theta_c^2}\right],\nonumber\\
\theta_-&=\frac12\left[(\beta-a)-\sqrt{(\beta-a)^2+4\theta_c^2}\right]. \label{eq:images}
\end{align}
For $a=0$ these reduce to the familiar point-lens result $\theta_\pm=\frac12[\beta\pm(\beta^2+4\theta_c^2)^{1/2}]$, with $\theta_c$ playing the role of the Einstein angle; the plasma and the effective-mass enhancement by $K$ therefore affect the image positions only through $\theta_c$. The angular separation is
\begin{align}
\Delta\theta&=\theta_+-\theta_-\nonumber\\
&=a+\frac12\left[\sqrt{(\beta+a)^2+4\theta_c^2}+\sqrt{(\beta-a)^2+4\theta_c^2}\right], \label{eq:separation}
\end{align}
which increases with both $K$ and $\eps$ and decreases negligibly with $Q$.

\subsection{Magnification}

The magnification of each image is the ratio of the solid angle subtended by the image to that of the unlensed source,
\begin{equation}
\mu_\pm=\frac{\theta_\pm}{\beta}\frac{d\theta_\pm}{d\beta},\qquad\frac{d\theta_\pm}{d\beta}=\frac12\left[1\pm\frac{\beta\pm a}{\sqrt{(\beta\pm a)^2+4\theta_c^2}}\right]. \label{eq:mu}
\end{equation}
For $a=0$, with $u=\beta/\theta_c$,
\begin{equation}
\mu_\pm=\frac{u^2+2}{2u\sqrt{u^2+4}}\pm\frac12,\quad\mu_{\rm tot}=|\mu_+|+|\mu_-|=\frac{u^2+2}{u\sqrt{u^2+4}}, \label{eq:mu_a0}
\end{equation}
and
\begin{equation}
\frac{|\mu_+|}{|\mu_-|}=\left[\frac{\sqrt{u^2+4}+u}{\sqrt{u^2+4}-u}\right]^2, \label{eq:flux_ratio}
\end{equation}
so that $|\mu_+|-|\mu_-|=1$ and the primary image is always the brighter one. The total magnification diverges as $u\to0$, the Einstein-ring limit, and tends to unity as $u\to\infty$. For $a=0$ the plasma and $K$ change these quantities only by rescaling $\theta_c$: at fixed $\beta$, a larger $\theta_c$ increases $\mu_{\rm tot}$ and brings $|\mu_\pm|/\mu_{\rm tot}$ closer to $1/2$.

Fig.~\ref{fig:mutot} shows $\mu_{\rm tot}$ as a function of $\beta/\thGR$. In panel (a), $a=0$ and the plasma raises $\mu_{\rm tot}$ at fixed $\beta$; at $\beta=\thGR$, $\mu_{\rm tot}$ increases from $1.34$ in vacuum to $1.51$ for $\eps=0.5$. In panel (b), $\eps=0$ and the deficit term raises $\mu_{\rm tot}$ more strongly, to $1.90$ for $a=0.5\,\thGR$ and $2.56$ for $a=\thGR$. Fig.~\ref{fig:mufrac} shows the fractional magnifications $|\mu_\pm|/\mu_{\rm tot}$. In all cases the primary image dominates for $\beta>0$, and both fractions tend to $1/2$ as $\beta\to0$. Increasing $\eps$ or $a$ slows the approach of the primary fraction to unity: at $\beta=\thGR$ it equals $0.873$ in vacuum, $0.831$ for $\eps=0.5$ and $0.805$ for $a=\thGR$. For $a\neq0$ the identity $|\mu_+|-|\mu_-|=1$ no longer holds; at $\beta=\thGR$ and $a=0.5\,\thGR$, $|\mu_+|-|\mu_-|=1.30$.

\section{Bounds from the Sloan Lens ACS Survey}\label{sec:slacs}

In this section we use observational data from the SLACS survey~\cite{bolton2006,koopmans2006,bolton2008} to bound the model parameters. Although numerous studies have constrained BH parameters using EHT observations, galaxy-scale strong lensing provides an independent avenue. The SLACS lenses are massive early-type galaxies with Einstein radii measured from HST/ACS F814W imaging and spectroscopic lens and source redshifts~\cite{bolton2008}.

Table~\ref{tab:slacs} lists the two systems used previously in Ref.~\cite{rahmatov2026}, with Einstein radii from the singular-isothermal-ellipsoid models of Ref.~\cite{bolton2008} and angular-diameter distances computed for a flat $\Lambda$CDM cosmology with $H_0=70$~km\,s$^{-1}$\,Mpc$^{-1}$ and $\Omega_m=0.3$. Two remarks on the use of these data are in order. First, the lens is a galaxy whose Einstein radius of several kiloparsecs encloses stars and dark matter, so the metric~\eqref{eq:metric} is used here as an effective description of the lens within the thin-lens approximation, not as a model of the central BH. Second, the lensing masses usually quoted for these systems, $M_E=\theta_E^2c^2D_LD_S/(4GD_{LS})$, are derived from $\theta_E$ assuming general relativity. Comparing a model prediction based on $M_E$ with the same $\theta_E$ is therefore circular and cannot constrain $K$, $Q$ or $\eps$. We use $M_E$ only to set the scale $r_s/R_E$ and derive bounds that do not rely on it.

\begin{table*}[h]
\caption{Selected SLACS lenses. $z_l$ and $z_s$ are the lens and source redshifts and $\theta_E$ the observed Einstein radius~\cite{bolton2008}. Distances assume flat $\Lambda$CDM with $H_0=70$~km\,s$^{-1}$\,Mpc$^{-1}$ and $\Omega_m=0.3$. $M_E$ is the Einstein mass implied by $\theta_E$ in general relativity, used only to evaluate $r_s/R_E$; $K_{\max}$ is the mass-independent bound of Eq.~\eqref{eq:Kmax}.}
\label{tab:slacs}
\begin{ruledtabular}
\begin{tabular}{lcccccccc}
Lens & $z_l$ & $z_s$ & $\theta_E$ (arcsec) & $D_L$ (Mpc) & $D_S$ (Mpc) & $D_{LS}$ (Mpc) & $M_E$ ($10^{11}M_\odot$) & $r_s/R_E$ \\
\hline
J0008$-$0004 & 0.440 & 1.192 & 1.16 & 1173 & 1709 & 939 & 3.53 & $5.1\times10^{-6}$\\
J0029$-$0055 & 0.227 & 0.931 & 0.96 & 750 & 1622 & 1145 & 1.20 & $3.3\times10^{-6}$
\end{tabular}
\end{ruledtabular}
\end{table*}

\begin{figure*}[t]
\centering
\includegraphics[width=0.92\textwidth]{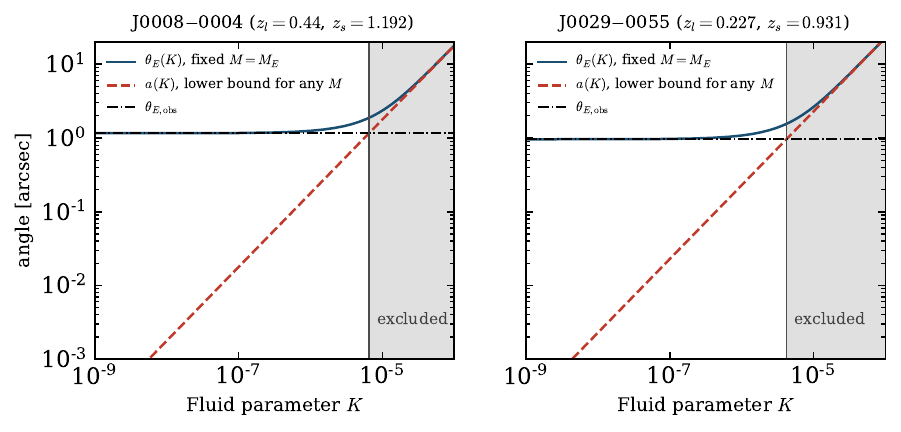}
\caption{Bound on the fluid parameter from the SLACS Einstein radii. Dashed: reduced deficit deflection $a(K)$, a lower limit on $\theta_E$ for any lens mass. Solid: $\theta_E(K)$ from Eq.~\eqref{eq:thetaE} for a fixed mass equal to the Einstein mass $M_E$ ($\eps=Q=0$). Dash-dotted: observed Einstein radius. The shaded region $K>K_{\max}$ [Eq.~\eqref{eq:Kmax}] is excluded independently of the mass.}
\label{fig:Kbound}
\end{figure*}

\begin{figure}[t]
\centering
\includegraphics[width=0.5\textwidth]{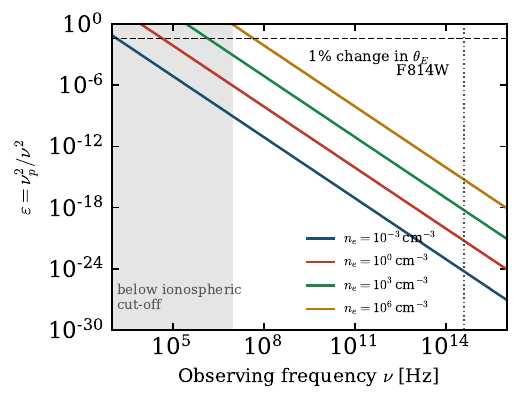}
\caption{Plasma parameter $\eps=\nu_p^2/\nu^2$ as a function of observing frequency for different electron densities. The dashed line marks the value giving a one per cent change in $\theta_E$; the dotted line marks the HST F814W band used for the SLACS Einstein radii; the shaded band indicates frequencies below the ionospheric cut-off.}
\label{fig:plasma}
\end{figure}

\subsection{Mass-independent bound on $K$}

From Eq.~\eqref{eq:thetaE}, $\theta_E>a$ for any positive lens mass. The observed Einstein angle therefore requires $a<\theta_{E,\rm obs}$, that is,
\begin{equation}
K<K_{\max}\equiv1-\left[1+\frac{D_S\,\theta_{E,\rm obs}}{\pi D_{LS}}\right]^{-2}\simeq\frac{2}{\pi}\frac{D_S}{D_{LS}}\,\theta_{E,\rm obs}. \label{eq:Kmax}
\end{equation}
This gives $K_{\max}=6.5\times10^{-6}$ for J0008$-$0004 and $K_{\max}=4.2\times10^{-6}$ for J0029$-$0055. The bound uses only $\theta_{E,\rm obs}$ and the redshifts, and it scales linearly with $\theta_{E,\rm obs}$, so a fractional uncertainty in the Einstein radius changes $K_{\max}$ by the same fraction. Fig.~\ref{fig:Kbound} illustrates it: the dashed curve is $a(K)$, below which the Einstein angle cannot fall, and the solid curve shows $\theta_E(K)$ for a fixed mass equal to $M_E$. Already for $K\sim10^{-7}$ the deficit term changes $\theta_E$ at fixed mass at the per cent level, so an independent mass estimate, for example from stellar dynamics combined with lensing~\cite{newman2026}, would tighten the bound considerably. The bound applies to $K$ as a parameter of the effective lens metric. Since the $w=-1/3$ Kiselev term extends to infinity, a consistent embedding in an expanding background would require a matching radius, which we leave for future work.

\subsection{Plasma}

At fixed mass the plasma changes the Einstein angle by the factor in Eq.~\eqref{eq:thetaE_a0}, $\theta_E/\theta_E(\eps=0)\simeq1+\eps/4$, which is degenerate with the lens mass. More importantly, $\eps$ is extremely small at the wavelengths of the SLACS data. With $\nu_p=8.98~{\rm kHz}\,(n_e/{\rm cm^{-3}})^{1/2}$ and $\nu\approx3.7\times10^{14}$~Hz for the F814W band, $\eps\approx6\times10^{-22}\,(n_e/{\rm cm^{-3}})$, so that even $n_e=10^6~{\rm cm^{-3}}$ along the entire path would give $\eps\approx6\times10^{-16}$. Fig.~\ref{fig:plasma} shows $\eps$ as a function of observing frequency. A one per cent change in $\theta_E$ requires $\eps\approx0.04$, that is $\nu\lesssim5\,\nu_p$, which corresponds to $46$~kHz for $n_e=1~{\rm cm^{-3}}$ and $1.4$~MHz for $n_e=10^3~{\rm cm^{-3}}$, below the ionospheric cut-off for ground-based observations. The SLACS Einstein radii therefore carry no information on $\eps$, and plasma effects of the type considered here can be probed only at low radio frequencies~\cite{bk2010}.

\section{Conclusion}\label{sec:conclusion}

In this work, we have investigated weak gravitational lensing by a charged BH embedded in an anisotropic matter distribution and surrounded by a homogeneous plasma, using the Hamiltonian formalism for photon propagation in a dispersive medium. The main results of this study are summarized as follows.
\begin{itemize}
\item The weak-field deflection angle consists of a term $\pi[(1-K)^{-1/2}-1]$, independent of the impact parameter, which arises from the solid-angle deficit; a $1/b_g$ term enhanced by $(1-K)^{-3/2}$ and by the plasma factor $[1+1/(1-(1-K)\eps)]/2$; and second-order terms in which the charge enters with a negative sign. The result agrees with direct numerical integration to a few parts in $10^4$ at $b=100\,r_s$.
\item The Einstein ring is enlarged by $K$, mainly through the deficit term, and by the plasma, with $\theta_E/\thGR\simeq1+\eps/4$ for $K=0$, while the charge reduces it by the fraction $(3\pi/16)Q^2r_s/R_E$.
\item In the absence of the deficit term, $K$ and $\eps$ affect the image positions and magnifications only through $\theta_c$; the primary image always dominates, with $|\mu_+|-|\mu_-|=1$, and at fixed source position the total magnification grows with $\eps$. The deficit term breaks this universality and increases the total magnification further.
\item The Einstein radii of J0008$-$0004 and J0029$-$0055 give the mass-independent bounds $K<6.5\times10^{-6}$ and $K<4.2\times10^{-6}$, because the Einstein angle cannot be smaller than the reduced deficit deflection.
\item In conclusion, plasma corrections to the Einstein angle are negligible at the optical wavelengths of the SLACS observations, with $\eps \simeq 6\times10^{-22}(n_e/\mathrm{cm}^{-3})$ in the F814W band. Consequently, the SLACS Einstein radii cannot meaningfully constrain $\eps$. Probing plasma-induced lensing effects may instead require low-frequency radio observations, where such effects are significantly enhanced.
\end{itemize}

These results show that galaxy-scale strong lensing places stringent limits on Kiselev-type matter with $w=-1/3$ (anisotropic matter) surrounding the lens, independently of the lens mass, whereas plasma effects require low-frequency radio observations~\cite{bk2010}. Future extensions of this analysis may incorporate independent dynamical masses~\cite{newman2026}, rotating or magnetised generalisations of the present spacetime, inhomogeneous plasma distributions, a consistent embedding of the deficit term in an expanding background, and larger samples from forthcoming strong-lensing surveys~\cite{sonnenfeld2024}.

\begin{acknowledgments}
The author RP acknowledges the research facilities at the IUCAA Centre for Astronomy Research and Development (ICARD), HNB Garhwal University. The author HN acknowledges financial support from the Anusandhan National Research Foundation (ANRF), through the Science and Engineering Research Board (SERB) Core Research Grant (Grant No.\ CRG/2023/008980).
\end{acknowledgments}

\section*{Data availability}
The SLACS data used in this work are publicly available in Ref.~\cite{bolton2008}. 


\end{document}